\documentclass[conference]{IEEEtran}
\IEEEoverridecommandlockouts
\usepackage{cite}
\usepackage{amsmath,amssymb,amsfonts}
\usepackage{algorithmic}
\usepackage{textcomp}
\usepackage{xcolor}
\usepackage{subfigure}
\usepackage{graphicx}
\usepackage{multirow}
\usepackage{makecell}
\usepackage{booktabs}
\usepackage{bm}
\usepackage{xcolor}
\usepackage{natbib}
\setcitestyle{numbers,square}

\def\BibTeX{{\rm B\kern-.05em{\sc i\kern-.025em b}\kern-.08em
    T\kern-.1667em\lower.7ex\hbox{E}\kern-.125emX}}

\renewcommand{\tablename}{\large TABLE}
\renewcommand{\thetable}{\large \Roman{table}}
\begin{document}

\title{Collaborative Automatic Modulation Classification via Deep Edge Inference for Hierarchical Cognitive Radio Networks}

\author{\IEEEauthorblockN{Chaowei He$^*$, Peihao Dong$^*$$^\dagger$, Fuhui Zhou$^*$, Qihui Wu$^*$}
	\IEEEauthorblockA{$^*$ College of Electronic and Information Engineering, Nanjing University of Aeronautics and Astronautics, Nanjing 211106, China\\
		$^\dagger$ National Mobile Communications Research Laboratory, Southeast University, Nanjing 211111, China\\
		Email: \{hcw0110, phdong\}@nuaa.edu.cn, zhoufuhui@ieee.org, wuqihui2014@sina.com}
}

\maketitle
\begin{abstract}
In hierarchical cognitive radio networks, edge or cloud servers utilize the data collected by edge devices for modulation classification, which, however, is faced with problems of the transmission overhead, data privacy, and computation load. In this article, an edge learning (EL) based framework jointly mobilizing the edge device and the edge server for intelligent co-inference is proposed to realize the collaborative automatic modulation classification (C-AMC) between them. A spectrum semantic compression neural network (SSCNet) with the lightweight structure is designed for the edge device to compress the collected raw data into a compact semantic message that is then sent to the edge server via the wireless channel. On the edge server side, a modulation classification neural network (MCNet) combining bidirectional long short-term memory (Bi-LSTM) and multi-head attention layers is elaborated to determine the modulation type from the noisy semantic message. By leveraging the computation resources of both the edge device and the edge server, high transmission overhead and risks of data privacy leakage are avoided. The simulation results verify the effectiveness of the proposed C-AMC framework, significantly reducing the model size and computational complexity.
\end{abstract}


\section{Introduction}

\IEEEPARstart{D}{espite} the significant advance, wireless networks still face critical challenges related to the spectrum scarcity and transmission security. Automatic modulation classification (AMC), initially developed for military use, now plays a vital role in addressing these issues. It enables receivers or monitors to identify signal modulation types with minimal prior knowledge, leading to its widespread application in civilian wireless networks including spectrum monitoring, smart modem development, and malicious attack detection \cite{O. A. Dobre}--\cite{Y. A. Eldemerdash}.

AMC algorithms fall into two main types: likelihood-based (LB) and feature-based (FB). LB methods compute the likelihoods of the candidate modulation schemes and choose the one with the maximum likelihood therein .However the need for the accurate channel information and high computational cost limit the practical use of this branch \cite{P. Panagiotou}, \cite{J. L. Xu}. FB methods, in contrast, analyze signal features like cyclic moments or wavelet-based attributes efficiently, offering near-optimal the classification performance \cite{W. A. Gardner}--\cite{L. Han}. However, traditional FB's reliance on manual feature extraction struggles with complex wireless conditions and the growing number of signal sources, highlighting the need for more scalable feature extraction techniques.

The revival of deep learning (DL) has revitalized FB approaches for AMC. Deep neural networks (DNNs) continue to emerge as powerful feature extractors and become the leading solution for modern AMC. In \cite{T. J. O'Shea_b},the research demonstrates the effectiveness of DL-based AMC in real-world conditions. In \cite{S. Peng}, convolutional neural networks (CNNs) have been specifically adapted for AMC to identify optimal data formats for received signals and improve classification accuracy. In \cite{F. Meng}, an end-to-end CNN architecture with the two-step training was developed for AMC. This DL-driven approach to AMC has inspired a surge of further research focused on enhancing feature extraction capabilities \cite{T. Huynh-The}--\cite{S. Chang}.

As wireless networks become more complex, there is an increasing need for efficient spectrum monitoring and security. This has led cognitive radio (CR) systems to evolve towards hierarchical structures that include edge devices, and edge or cloud servers \cite{P. Dong}. However, directly applying traditional deep learning models for AMC  to such structures is challenging. In \cite{S. Rajendran}, the research has explored using long short-term memory (LSTM) networks for AMC in distributed sensors. While this method is practical, it places a strain on the constrained resources of devices  and risks data exposure. Edge learning offers a solution by partitioning DNN models between devices and servers, minimizing data transfer, enhancing privacy, and distributing computational load \cite{W. Xu}. Implementing DNN model splitting on residual networks can lead to high transmission volumes and depend heavily on sufficient in-phase/quadrature (I/Q) samples, which puts the additional load on edge devices \cite{J. Park}.

To address the mentioned-above problems, in this article, we propose an EL-based collaborative automatic modulation classification (C-AMC) framework consisted of a spectrum semantics compression neural network (SSCNet) and a modulation classification neural network (MCNet) deployed at the edge device and edge server respectively. The main novelty and contribution can be summarized as follows:

\begin{itemize}
\item[1)] The C-AMC architecture is proposed to light the computing load of the edge device, reduce the transmission overhead, and protect the privacy of the original data, which meets the growing demand for secure and low-latency communication in modern wireless networks.
\item[2)]SSCNet is designed to be lightweight and easily deployable on edge devices. Meanwhile, MCNet incorporates bidirectional LSTM and attention mechanisms, enabling it to thoroughly extract signal features for high-precision classification. This combination of SSCNet and MCNet in the C-AMC framework allows for a balance between efficient data processing and robust classification capabilities. 
\item[3)] The simulation results demonstrate the superior peformance of the proposed C-AMC framwork with the low complexity and transmission overhead. The impacts of some key factors, the compression rate, signal-to-noise ratios of the sensing channel and transmission channel on the classification accuracy are also revealed, providing valuable insights for practical applications.
\end{itemize}
    
The rest of this paper is organized as follows. Section II introduces the basic system model, based on which the EL-based C-AMC framework is developed in Section III. Simulation results are presented in Section IV. Finally, we summarize the content of this paper in Section V.
    
\section{System Model}

In this section, we detail the hierarchical CR network model by treating the modulation classification as the cognitive task.

\begin{figure}[t]
	\centering
	\includegraphics[width=3.4in]{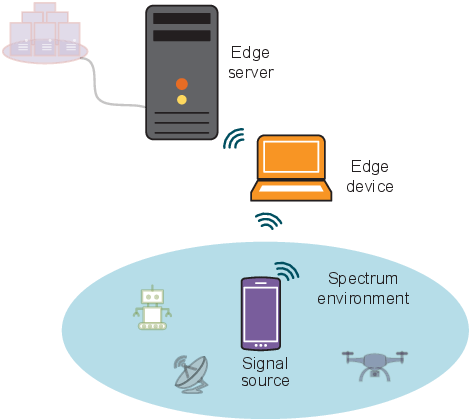}
	\caption{Hierarchical CR network model.}\label{HCR_model}
\end{figure}

Fig.~\ref{HCR_model} shows a hierarchical CR network, comprising an edge device and an edge server, designed to recognize the modulation type of signals in the spectrum environment. This cognitive process includes two phases: the sensing of data by edge devices and the subsequent transmission of the data to the edge server, accompanied by the requisite signal processing steps.

\begin{figure*}[t]
	\centering
	\includegraphics[width=4.9in]{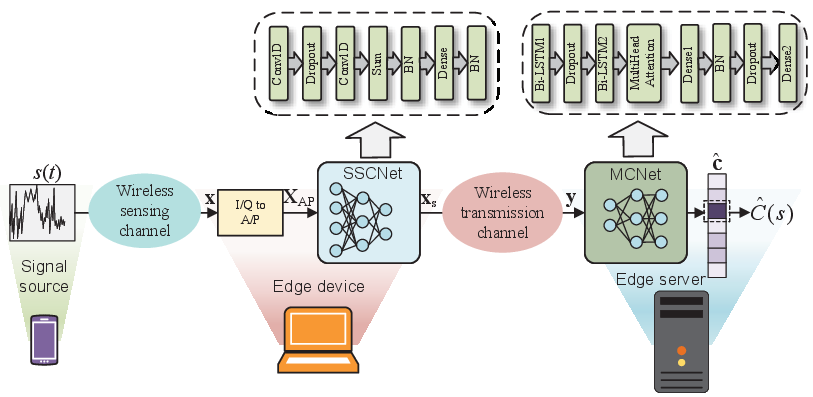}
	\caption{EL-based C-AMC framework.}\label{C_AMC}
\end{figure*}

\emph{Phase I (Data Sensing at the Edge Device):} The edge device continuously probes the source signal $s(t)$ within $L$  sampling time instants, and the signal that is received can be described as
\begin{eqnarray}
	\label{eqn_xl}
	x[l] = h[l]e^{-j2\pi(\nu lT_{\mathrm{s}}+\theta)}s[l] + z[l],\quad l=1,\ldots,L,
\end{eqnarray}
where $h[l]$, $s[l]$, and $z[l]$ respectively symbolize the channel gain, modulated source signal, and additive white Gaussian noise (AWGN) at the $l$th sampling time instant. Additionally, $T_{\mathrm{s}}$, $\nu$, and $\theta$ stand for the sampling interval, frequency offset, and phase shift, correspondingly. The modulation type of the source signal, $C(s)$, is selected from a set of potential types $\mathcal{M}$, which contains 
$M$ elements. Let $\mathbf{x}=[x[1],\ldots,x[L]]$ represent the vectorized form of the received signal. This vector $\mathbf{x}$ undergoes processing within the edge device to produce an $N$-dimensional vector expressed as
\begin{eqnarray}
	\label{eqn_xs_vec}
	\mathbf{x}_{\textrm{s}}=\phi(\mathbf{x}),
\end{eqnarray}
where $\phi(\cdot)$ symbolizes the general mapping function used in signal processing. In particular, $\mathbf{x}_{\textrm{s}}$ will output the classification result when $\phi(\cdot)$ acts as the mapping function for modulation classification, whereas $\mathbf{x}_{\textrm{s}}=\mathbf{x}$ applies when $\phi(\cdot)$ serves as the identity mapping.

\emph{Phase II (Information Transmission from the Edge Device to the Edge Server):} In this phase, the edge device send $\mathbf{x}_{\textrm{s}}$ to the edge server through a certain air interface. The received signal after equalization by the edge server is represented as

\begin{eqnarray}
	\label{eqn_y_elem_wise}
	y_{i}=\bar{x}_{\textrm{s},i}+w_i,\quad i=1,\ldots,N,
\end{eqnarray}
where $\bar{x}_{\textrm{s},i}$ represents the $i$th element of $\mathbf{x}_{\textrm{s}}$, and $w_i$ signifies the associated effective noise that combines both interference and AWGN. Following this, the edge server performs the operation $\varphi(\cdot)$, on $\mathbf{y}=[y_{1},\ldots,y_{N}]$ to  identify the modulation type of the originating signal. Assuming a scenario without noise, the result is given as
\begin{eqnarray}
	\label{eqn_c}
	\hat{\mathbf{c}}=\varphi(\mathbf{y})=\varphi(\phi(\mathbf{x})),
\end{eqnarray}
where $\hat{\mathbf{c}}\in \mathbb{R}^{M}$ is a vector of probabilities that points to the most probable modulation type of $s(t)$ determined by the largest value's index within the vector. As illustrated in (\ref{eqn_c}), the functions $\phi(\cdot)$ and $\varphi(\cdot)$ work together to complete the task of modulation classification in a synergistic fashion. In the scenario described above, $\varphi(\cdot)$ can act as the identity mapping when $\phi(\cdot)$ performs modulation classification mapping, and the reverse is also true. However, this approach can increase the computational demands on the edge device or lead to high transmission costs, alongside the potential risk of compromising sensitive information. Therefore, we aim to address these issues by further developing $\phi(\cdot)$ and $\varphi(\cdot)$ in the subsequent sections.

\section{EL-Based C-AMC Framework}

This section introduces the construction of the EL-empowered C-AMC framework. We begin with an overview of the framework, then delve into the specifics of its two DL component models.

\subsection{Overview of Framework}

Inspired by EL, the C-AMC framework effectively balances the computational demands of modulation classification between the edge device and the edge server by specifically deploying SSCNet and MCNet to each, as shown in Fig. 2. The task of SSCNet is to compress the sensed data into a compact spectral semantic embedding by extracting features related to its modulation type. On the receiving end of the wireless channel, MCNet leverages this noisy semantic embedding to predict the source signal’s modulation type. This strategy allows a portion of the MC computational workload to be shifted from the resource-constrained edge device to the more capable edge server. 

A compact spectral semantic embedding represents high-dimensional spectral data in a lower-dimensional, meaningful form. This approach maps high-dimensional data into a lower-dimensional space while preserving meaningful relationships and features. In this context, semantic embedding retains information about modulation types. Semantic embedding reduces complexity and speeds up signal analysis and classification. Additionally, the compact nature of the semantic embedding not only minimizes transmission overhead but also poses a challenge for potential interceptors attempting to decode it in the wireless channel.

Before SSCNet processes it at the edge device, the sensed data $\mathbf{x}$ in the I/Q form is transformed into the amplitude/phase (A/P) form as
\begin{eqnarray}
	\label{eqn_AP}
	\mathbf{X}_{\textrm{AP}}=\mathcal{P}(\mathbf{x})=\left[\begin{array}{cc}|x[1]|,\ldots,|x[L]|\\ \theta(x[1]),\ldots,\theta(x[L])\end{array}\right]^{T},
\end{eqnarray}
where $|x|$ represents the amplitude of $x$, and $\theta(x)=\mathrm{arctan}\frac{\mathrm{Im}(x)}{\mathrm{Re}(x)}$ signifies the phase of $x$. Following this, $\mathbf{x}_{\textrm{AP}}$ is processed by SSCNet to generate the spectrum semantic embedding. The compression rate $r=\frac{2L}{N}$, where the input  $\mathbf{x}_{\textrm{AP}}\in \mathbb{R}^{L \times 2}$ and $N$ is the dimension of the semantic embedding. A higher compression rate indicates better compression, but it does not necessarily lead to better classification performance. The compression rate is achieved in this manner,
\begin{eqnarray}
	\label{eqn_xs_vec_SSCNet}
	\mathbf{x}_{\textrm{s}}=f(\mathbf{X}_{\textrm{AP}};\boldsymbol{\Theta}),
\end{eqnarray}
where $f(\cdot)$ represents the mapping function of SSCNet, which is parameterized by the set of weights $\boldsymbol{\Theta}$. Therefore, the general mapping function $\phi(\cdot)$ can be expressed as the composition of $\mathcal{P}(\cdot)$ and $f(\cdot)$, i.e., $\phi=f\circ \mathcal{P}$.

At the edge server, the semantic embedding with noise $\mathbf{y}=\mathbf{x}_{\textrm{s}}+\mathbf{w}$ , where  $\mathbf{w}=[w_{1},\ldots,w_{N}]$ is processed by MCNet to predict the modulation type, described as
\begin{eqnarray}
	\label{eqn_c_MCNet}
	\hat{\mathbf{c}}=h(\mathbf{y};\boldsymbol{\Phi})=h\left(f\left(\mathcal{P}(\mathbf{x});\boldsymbol{\Theta}\right)+\mathbf{w};\boldsymbol{\Phi}\right),
\end{eqnarray}
where $h(\cdot)$ refers to the mapping function of MCNet, which is parameterized by its weight set $\boldsymbol{\Phi}$. Therefore, the general mapping function $\varphi(\cdot)$ is instantiated as $h(\cdot)$, i.e., $\varphi=h$. Given that $\mathcal{P}(\cdot)$ represents a fixed manual operation, the design of $\phi(\cdot)$ and $\varphi(\cdot)$ effectively boils down to the architectural design of SSCNet and MCNet, represented by $f(\cdot)$ and $h(\cdot)$, respectively.

\subsection{SSCNet}

In the C-AMC framework, the role of SSCNet is to use a lightweight architecture to compress high-dimensional sensory data into compact semantic embeddings to adapt to edge devices with limited resources. To fully compress the sensing data, we use a combination of two one-dimensional convolutions (Conv1D), a lightweight solution for handling temporal correlations.

Fig.~\ref{C_AMC} shows the architecture of SSCNet. The first Conv1D layer using $64$ kernels with the length of $8$ and rectified linear unit (ReLU) 
activation function is utilized first to filter the input $\mathbf{x}_{\textrm{AP}}\in \mathbb{R}^{L \times 2}$, in order to extract the local temporal correlation therein, after which a dropout layer with the dropout rate of $0.5$ is added to prevent overfitting and to improve the robustness of neurons. Then, the $L \times 64$ feature map is processed by the second Conv1D layer to further extract the required spectral features. The second Conv1D layer has $32$ kernels of length $8$ also using the rectified linear unit (ReLU) activation function. Through the second Conv1D layer, an $L \times 32$ feature map is generated.
After that this feature map is summed in the column-wise manner to yield a $1 \times 32$ feature vector, which then undergoes the batch normalization (BN) layer.The following dense layer compresses the feature vector and the final BN layer outputs the $N$-dimensional real-valued specturm semantic embedding, $\mathbf{x}_{\textrm{s}}$, with $N \ll 2L$. The dense layers adopt scaled exponential linear unit (SELU) activation function to ensure the neuron activity regardless of the sign of the input, 
that is,
\begin{eqnarray}
	\label{eqn_SELU}
	\mathrm{SELU}(x)=\lambda \begin{cases}x, & x>0 \\ \alpha\left(e^x-1\right), & x \leq 0\end{cases}
\end{eqnarray}
with $\alpha>0$ and the scaling factor $\lambda>1$.

\subsection{MCNet}

MCNet should be designed to coordinate with SSCNet so that they can be integrated across the wireless channel to output the predicted modulation type at the edge server. The Bi-LSTM structure is utilized to uncover the correlation hidden in the noisy semantic embedding $\mathbf{y}$ as well as to remove the noise imposed by the wireless channel. Compared with SSCNet, more Bi-LSTM layers can be applied in MCNet to fully extract the desired feature thanks to the much lower dimension of $\mathbf{y}$.

Fig.~\ref{C_AMC} shows the architecture of MCNet. Bi-LSTM1 layer includes $N$ units, each of which transforms the corresponding element of $\mathbf{y}$ into a $128$-dimensional vector, yielding the $N \times 128$ feature map. After Bi-LSTM1 layer, a dropout layer with the dropout rate of $0.5$ is appended. Bi-LSTM2 layer also includes $N$ units while the sizes of input and output vectors of each unit both are $128$. Then, the output of the Bi-LSTM2 layer is reshaped to $1 \times 128$ via the Reshape layer and sent to the multi-head attention layer, which sets an $8$ attention head. The key dimension of each head is $128$, and the query, key and value are all derived from the output of Bi-LSTM2. The multi-head attention layer aims to efficiently extract and synthesize features from the input data for easy classification. Dense1 layer enlarges the size of the feature vector to $256$, after which a BN layer and a dropout layer with the dropout rate of $0.5$ are appended successively. Finally, Dense2 layer outputs the probability vector $\hat{\mathbf{c}}$ by using Softmax activation function. SELU is applied for the hidden layers Dense1.
Finally, Dense1 layer outputs the probability vector $\hat{\mathbf{c}}$ by using Softmax activation function. 

\begin{figure}[t]
	\centering
	\includegraphics[width=2.7in]{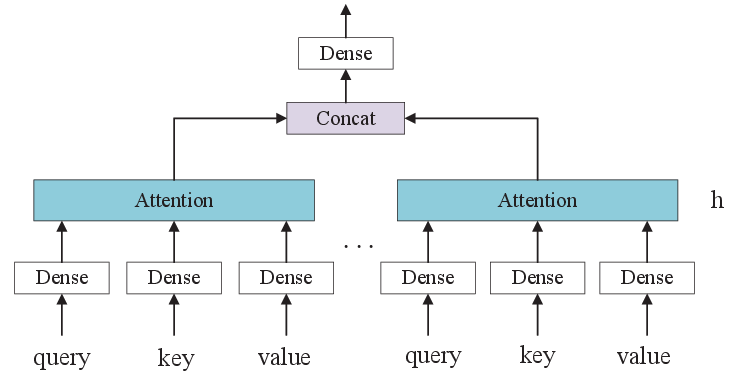}
	\caption{Multi-Head Attention.}\label{Multihead-attention}
\end{figure}

Fig.~\ref{Multihead-attention} shows the workflow of the multi-head attention mechanism. The multi-head attention layer receives the input $\mathbf{x}$ from the Bi-LSTM2 layer. The dimension of $\mathbf{x}$ is \(1 \times 128\). For each head in the attention layer, the input is processed through a dense layer with 128 units to construct query ($\mathbf{Q}$), key ($\mathbf{K}$), and value ($\mathbf{V}$) vectors, $\mathbf{Q}$ and $\mathbf{K}$ have the same dimension \(d_k\). Therefore, the output of the scaled dot product Attention can be written as
\begin{eqnarray}
  \mathit{Attention}(\mathbf{Q},\mathbf{K}, \mathbf{V}) = \mathit{softmax}\left(\frac{\mathbf{Q}\mathbf{K}^T}{\sqrt{d_k}}\right)\mathbf{V}.
\end{eqnarray}

In this equation, $\mathbf{K}^T$ means the transpose of $\mathbf{K}$, the output vector is the reweighted value $\mathbf{V}$, the weight assigned to each value is calculated by $\mathbf{K}$ and $\mathbf{Q}$.The multi-head mechanism divides $\mathbf{Q}$, $\mathbf{K}$ and
$\mathbf{V}$ into $h$ parts, undertakes scaled dot-product attention $h$ times and finally concatenates all $h$ outputs, which can be denoted by
\begin{align}
	&\mathit{MultiHead}(\mathbf{Q},\mathbf{K},\mathbf{V}) = \mathit{Concat}(\mathit{head}_1, \mathit{head}_2, \ldots, \mathit{head}_h)\mathbf{W}^O, \nonumber
\end{align}
\begin{equation}
	\mathit{head}_i = \mathit{Attention}(\mathbf{x}\mathbf{W}^Q_i, \mathbf{x}\mathbf{W}^K_i, \mathbf{x}\mathbf{W}^V_i),
\end{equation}
where $\mathbf{W}^Q_i$, $\mathbf{W}^K_i$, $\mathbf{W}^V_i \in \mathbb{R}^{d_{\text{model}} \times d_k}$ denote the weight matrices used by the $i$-th head to generate the query, key, and value matrices.$\mathbf{W}^O \in \mathbb{R}^{h d_v \times d_{\text{model}}}$ denotes the weight matrices used to linearly transform the outputs of all heads in the multi-head attention mechanism to produce a unified output representation. The output of multi-head attention the concatenation of all heads, subsequently undergoing a linear transformation via an additional dense layer, and the output dimension is \(1 \times 128\).

\section{Simulation Results}

\subsection{Dataset}
In this study, we use the widely recognized RadioML2016.10A dataset as the basis for performance verification of the modulation identification framework. The dataset contains complex-valued I/Q samples for 11 different modulation modes. 

\subsection{Simulation Settings}

The output layer of our network is designed with 11 units to match the needs of 11 different modulation types in the dataset. Each modulation signal consists of 512 I/Q samples. We use cross-entropy loss function and Adam optimizer to help model learning. In order to optimize the training process, we initially set the learning rate to 0.001. In addition, our training process adopts the mini-batch gradient descent method, and the batch size is set to 200 to process the data efficiently. If no loss reduction is observed for 30 consecutive epochs, training is stopped early, which allows efficient use of computing resources.
\begin{table*}[t]
	\caption{\large Comparison of Prediction Accuracy, Parameters, FLOPs, and Inference Time}
	\label{tab:performance_comparison}
	\resizebox{\textwidth}{!}{
		\begin{tabular}{cccccc}
			\toprule
			Model & $P_{\mathrm{acc}}$ & Parameters & FLOPs & Inference Time & Compression Rate \\
			\midrule	
			C-AMC(N=64)  & \textbf{0.9322} & 20.00 K  &\textbf{17.9 M}  & \textbf{0.022 ms}  & \textbf{16}\\
			LSTMNet-DC       & 0.924 & 20.5 K & 21.3 M  & 0.065 ms  & -\\
			SSCNet-DC          & 0.916 & \textbf{18.00 K} & \textbf{17.9M}& \textbf{0.022 ms} & -\\
			SplitAMC(2,2) & 0.839 & 679.7 M & 150 M  & 0.133 ms  & 0.125\\
			SplitAMC(1,3) & 0.834 & 152.2 K & 82.5 M  & 0.069 ms  & 0.0625\\
			\bottomrule
		\end{tabular}
	}
\end{table*}

\begin{figure}[t]
	\centering
	\includegraphics[width=3.4in]{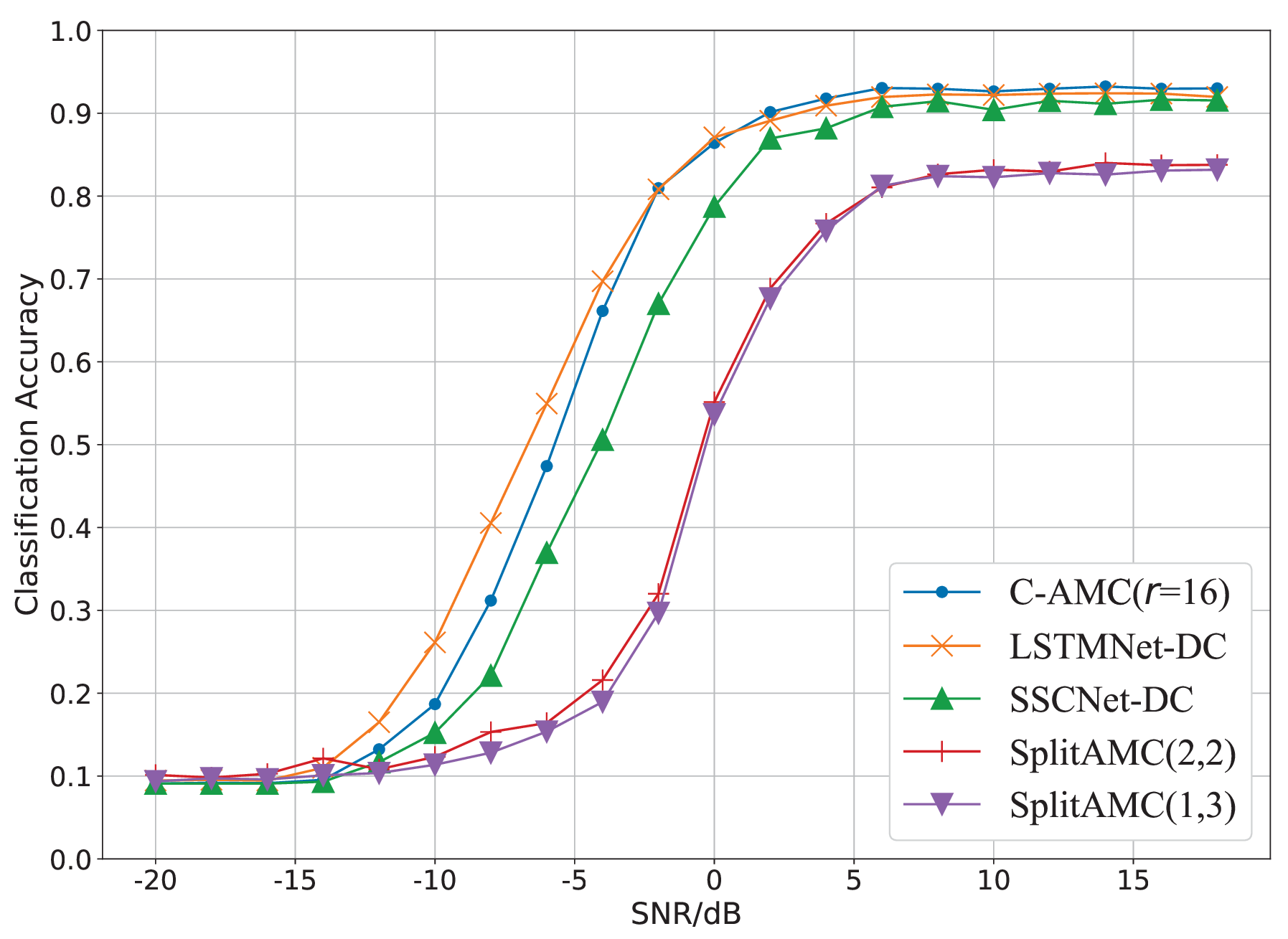}
	\caption{Performance of C-AMC,LSTMNet-DC,SSCNet-DC and SplitAMC on RadioML2016.10A dataset.}\label{C-AMC_compare_with_SSCNet_LSTM_SplitAMC}
\end{figure}

\begin{figure}[t]
	\centering
	\includegraphics[width=3.4in]{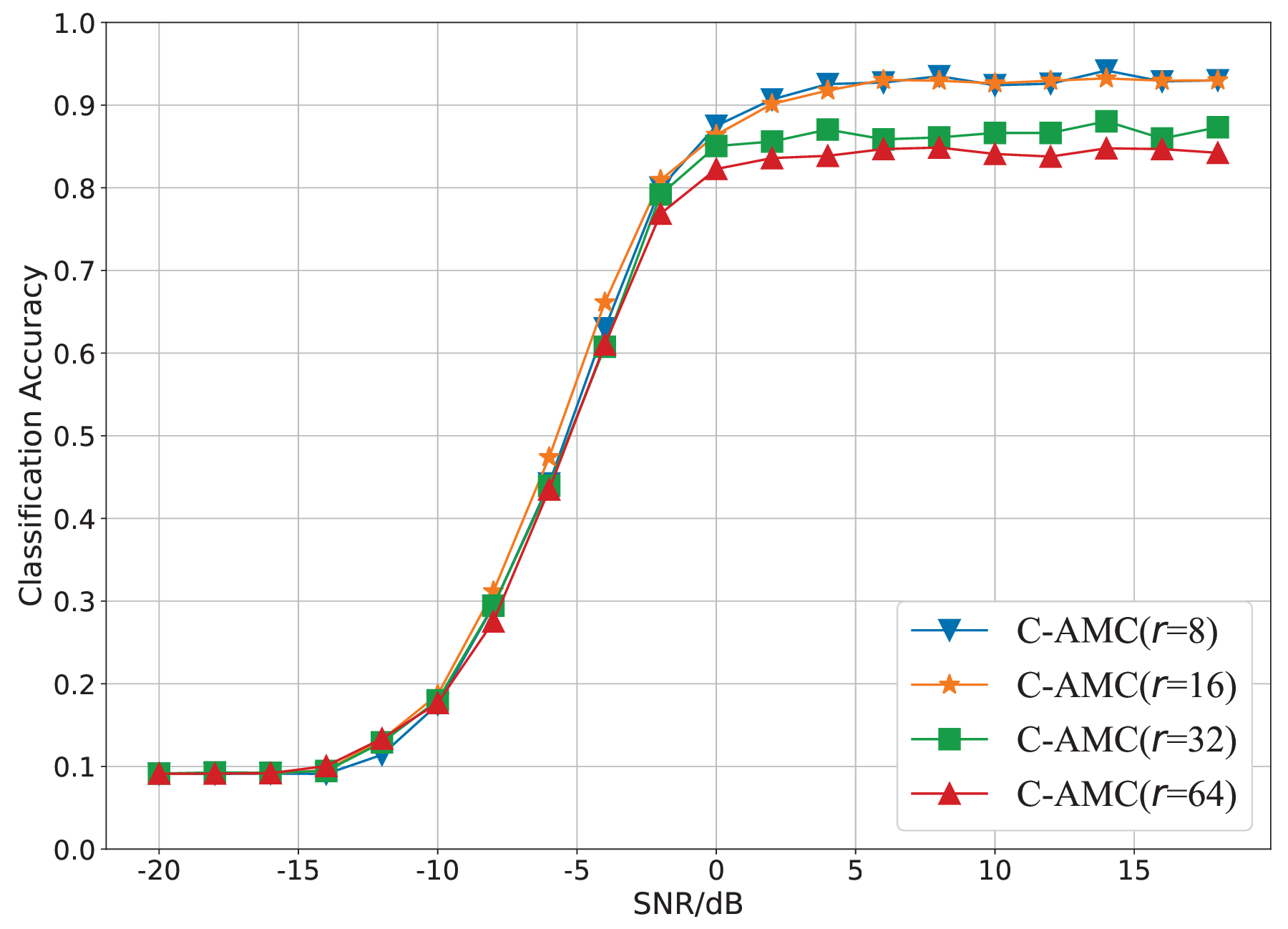}
	\caption{Performance of C-AMC on RadioML2016.10A dataset under different compression rate $r$.}\label{C-AMC compare}
\end{figure}

\subsection{The Performance of the C-AMC}

In Fig.~\ref{C-AMC_compare_with_SSCNet_LSTM_SplitAMC}, we compare the performance of different models on the RadioML2016.10A dataset. Among them, the "SSCNet-DC" curve shows the change in accuracy when adding a fully connected layer after SSCNet for direct classification. The curve marked "LSTMNet-DC" shows the model accuracy using a two-layer, 40-unit LSTM structure per layer.
In this comparison, the performance of the SplitAMC(2,2) and SplitAMC(1,3) models is relatively low, with the highest accuracy reaching only 83.7\% and 83.1\% respectively. For high SNR situations, the classification accuracy of C-AMC ($r$=16) is 1\% to 3\% higher than that of SSCNet-DC. This result highlights that the model with MCNet has significant performance advantages in the AMC task compared to the model without MCNet. In addition, we also notice that under high SNR conditions, the performance of the proposed C-AMC ($r$=16) framework exceeded the LSTMNet-DC model and is far better than the SplitAMC model, which further proves the excellence of the C-AMC model performance.

Table I provides a detailed comparison of C-AMC ($r$=16), LSTMNet-DC, SSCNet-DC and SplitAMC in terms of prediction accuracy, number of parameters, number of FLOPs, inference time, and compression rate on edge device. This comparison underscores the significant advantages of the C-AMC model in enhancing performance, reducing model size, and accelerating inference speeds.

Fig.~\ref{C-AMC compare} shows the performance of C-AMC on the RadioML2016.10A dataset under different compression rate $r$. For the C-AMC model, as the $r$ becomes bigger, the classification accuracy of the model gradually decreases. When $r$=8, the C-AMC model has the best performance, with an average classification accuracy of 91.94\% in the SNR range of 0 dB~18 dB. When SNR = 14dB, the model reaches peak performance and the classification accuracy is 94.18\%. When $r$=32, the model classification accuracy drops significantly, and the highest classification accuracy only reaches 88.04\%. This result clearly demonstrates the critical impact of the compression rate on the performance of the C-AMC model.

\begin{figure}[t]
	\centering
	\subfigure[]{
		\includegraphics[width=0.45\linewidth]{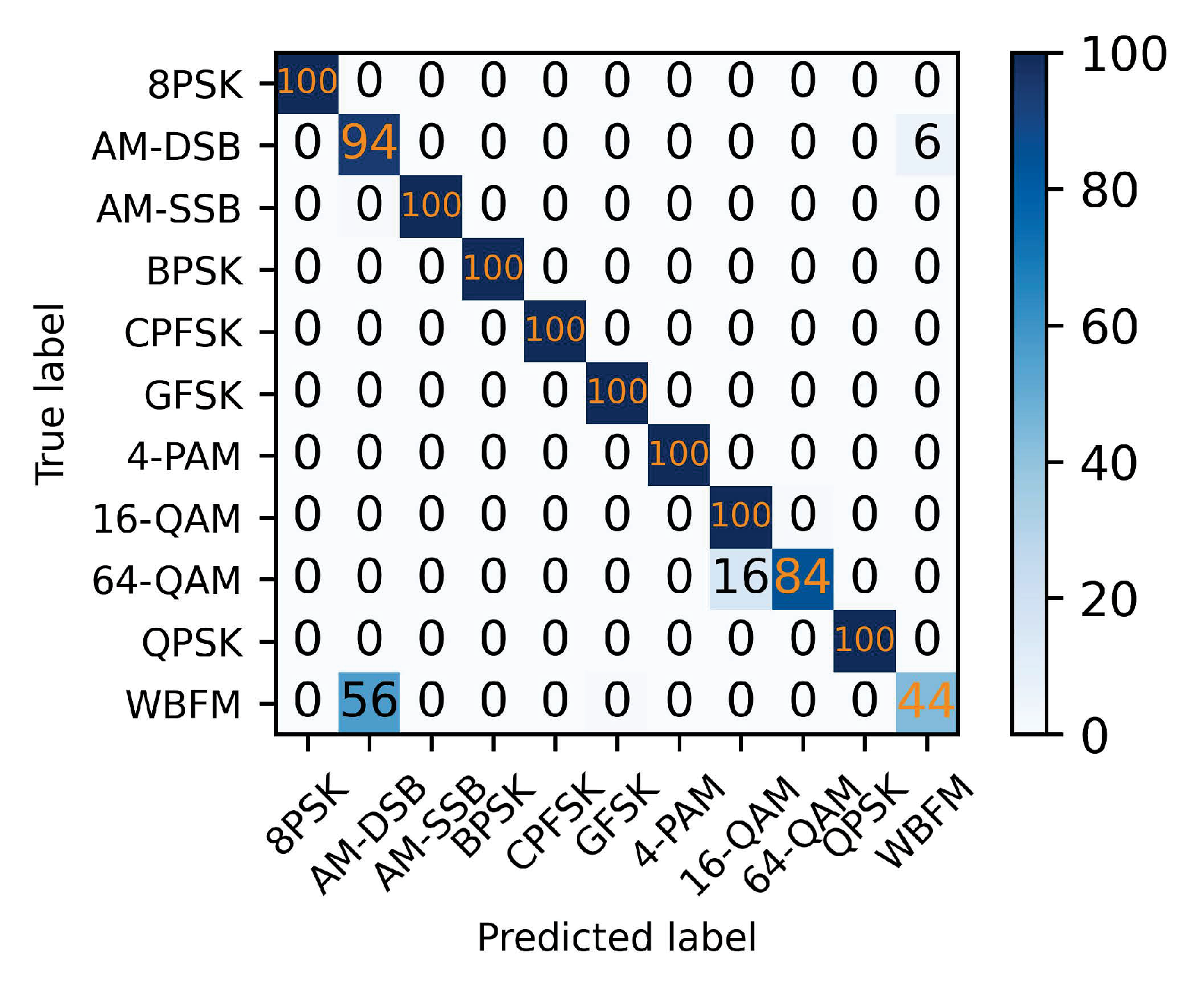}
		\label{SNR=16}
	}
	\subfigure[]{
		\includegraphics[width=0.45\linewidth]{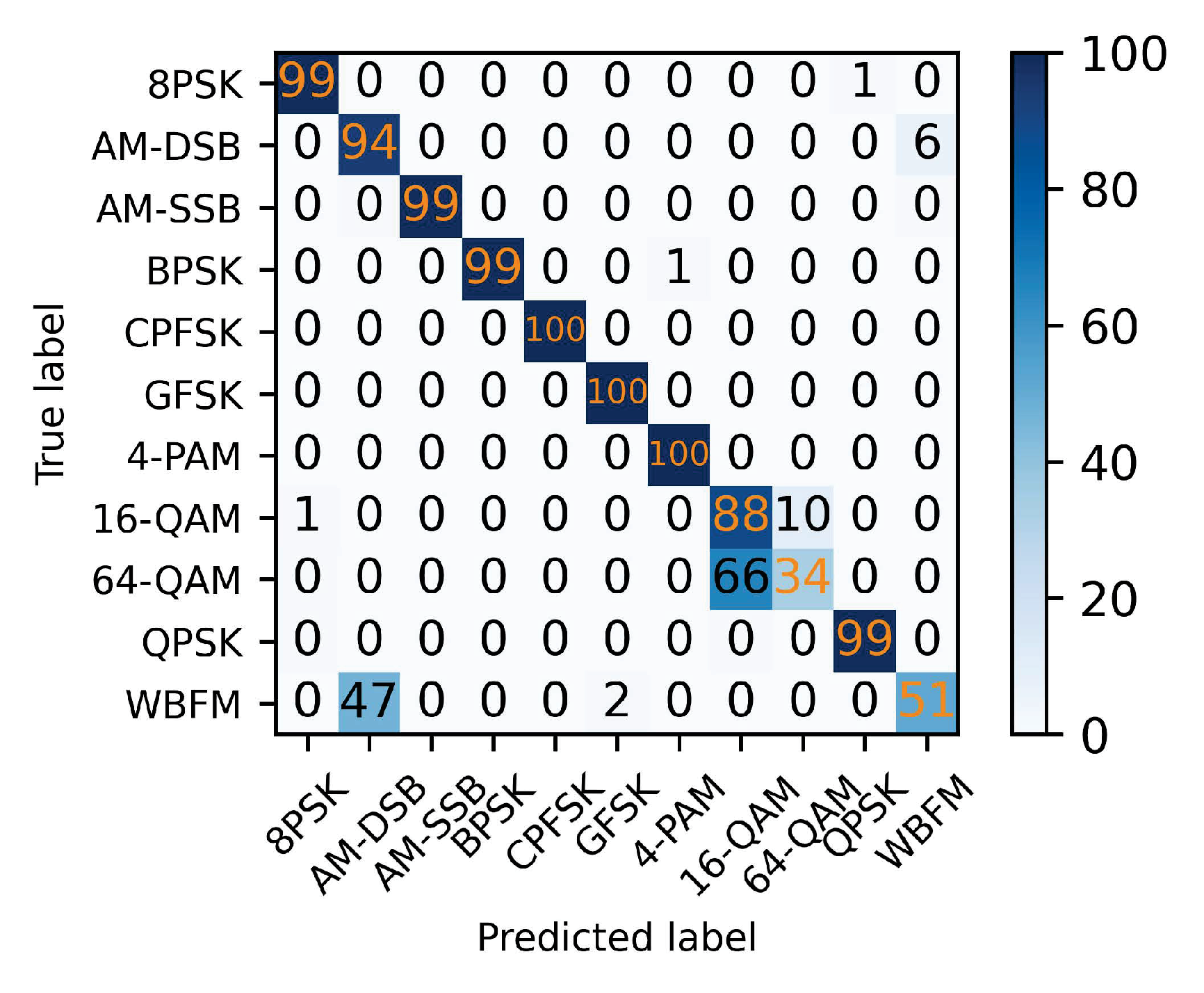}
		\label{SNR=0}
	}
	\vspace{0.5cm}
	
	\caption{The confusion matrices of the C-AMC($r$=16) for different SNRs: (a) 16 dB, (b) 0 dB.}
	\label{confusion matrices}
\end{figure}

Fig.~\ref{confusion matrices} displays the confusion matrix for C-AMC($r$=16) at SNRs of 16dB and 0dB. The performance of the proposed C-AMC($r$=16) demonstrates highly effective results at higher and medium SNR levels. The confusion primarily stems from two factors. Firstly, AM-DSB and WBFM are susceptible to misclassification because the spectrum of AM-DSB exhibits two symmetrical sidebands around a center frequency, lacking an accompanying carrier signal. Although WBFM modulation generates a signal with a broader bandwidth, its spectral characteristics can sometimes resemble those of AM-DSB, especially at lower SNRs.
Secondly, QAM16 and QAM64 are prone to misidentification due to overlapping constellation points and similar time-spectrum diagrams.
\begin{figure}[t]
	\centering
	\includegraphics[width=2.7in]{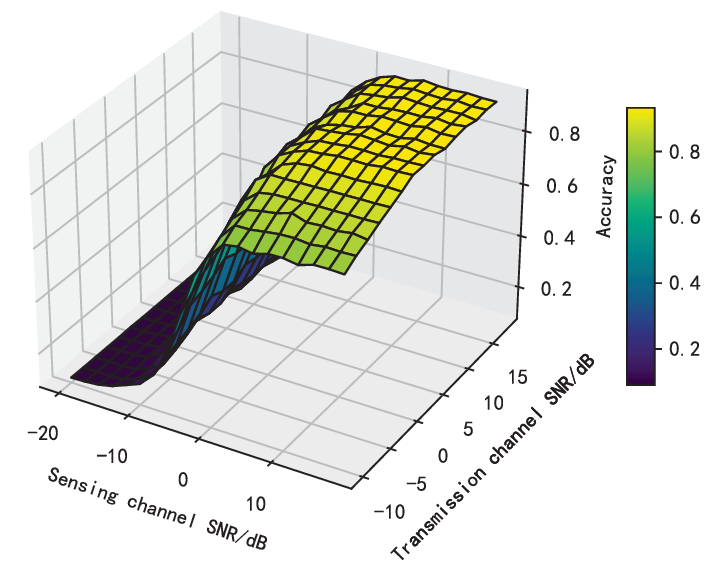}
	\caption{Performance of C-AMC ($r$=16) on different sensing and transmission channel SNRs.}\label{C-AMC three-dimensional map}
\end{figure}

Finally, Fig.~\ref{C-AMC three-dimensional map} shows the classification accuracy of C-AMC ($r$=16) across various the sensing and transmission channel SNRs. Notably, when the transmission channel SNRs in  the range of 2dB to 18dB, C-AMC ($r$=16) exhibited exceptional performance in  the high sensing  channel SNRs. Even amidst moderate transmission channel SNRs , ranging from -10dB to 0dB, the classification accuracy of C-AMC ($r$=16) experienced only a marginal decline, yet still maintained a commendable level of accuracy in  the high sensing  channel SNRs. This underscores the robustness of C-AMC ($r$=16) in effectively mitigating noise interference, thereby yielding superior results. Consequently, it can be inferred that C-AMC ($r$=16) possesses significant anti-noise capabilities in the context of transmission channel noise.

\section{Conclusion}

In this paper, we propose a novel AMC framework. The time-frequency representation of the received signal is used as input to the proposed C-AMC network. Different from existing modulation identification methods, the proposed collaborative automatic modulation classification framework based on edge learning effectively copes with the transmission overhead, data privacy and computational load issues in hierarchical cognitive wireless networks. Through the collaborative work of edge devices and servers, we designed SSCNet and MCNet, which not only achieve efficient signal processing, but also ensure data privacy, while significantly reducing the size and computational complexity of the model. In experiments, the proposed framework achieves the high calssification accuracy on the RadioML2016.10A dataset. This framework demonstrates the great potential of edge learning in smart wireless communication systems and provides a valuable reference for future research and practice.

\end{document}